# A Fundamental Scale for Acceleration from the Holographic Principle

Scott Funkhouser
*Occidental College, the Joint Sciences Department of the Claremont Colleges*

From the Eddington-Weinberg relationship, which may be explained by the holographic principle in a flat Universe, it follows that the characteristic gravitational acceleration $a_N$ associated with the nucleon and its Compton wavelength is of order the Hubble acceleration $H_0 c$ in this epoch. A natural scaling for the cosmological constant is obtained from this acceleration term. It also happens that the critical acceleration $a_0$ associated with the MOND theory is of order $a_N$.

Eddington first identified the perplexing occurrences of large pure numbers of order $10^{40}$ that may be obtained from the parameters of the Universe [1]. Many of those coincidences among large numbers may be combined to form the expression known as the Eddington-Weinberg relation

$$h^2 H_0 \sim G c m_N^3, \qquad (1)$$

where $h$ is the Planck's constant, $H_0$ is the current value of the Hubble parameter, $G$ is Newton's constant of gravitation and $m_N$ is the mass scale of the nucleon [2]. It has been found that Eq. (1) may be explained in terms of the cosmic coincidence and the holographic principle in a flat Universe [3]. According to the holographic principle the number of available states $N$ in the observable Universe is roughly the number of Planck areas contained on the surface of a sphere whose radius is the current Hubble radius $R_0$

$$N \sim \frac{R_0^2}{l_P^2} \sim 10^{120} \qquad (2)$$

where $l_P$ is the Planck length. If nucleons may be represented as quanta of an effective field within the spatial volume of a sphere whose radius is of order the Hubble radius then the number of degrees of freedom $N_N$ available to them may be given by

$$N_N \sim \frac{R_0^3}{\lambda_N^3} \qquad (3)$$

where $\lambda_N$ is the Compton wavelength of the nucleon. $N_N$ also happens to be of order near $10^{120}$ and this correspondence is consistent with the limitations imposed by the holographic principle [3].

Eq. (1) may be stated in a more physically suggestive form in terms of the Compton wavelength of the nucleon

$$H_0 c \sim \frac{G m_N}{\lambda_N^2}. \qquad (4)$$

The term on the right in Eq. (4) is the gravitational acceleration associated with the nucleon mass and its Compton wavelength and the term on the left is the current value of the Hubble acceleration. A characteristic time $t_N$ may be stated from the characteristic acceleration $a_N \equiv G m_N / \lambda_N^2$

$$t_N \equiv \frac{c}{a_N} \qquad (5)$$

and this time happens to be of order the age of the Universe.

The existence of a cosmological constant $\Lambda$ may explain the apparent accelerated expansion of the Universe. Due to the cosmic coincidence $G\rho_0 \sim \Lambda$ where $\rho_0$ is the current density of matter, the Friedmann equations lead to $H_0 \sim \sqrt{\Lambda}$. With this substitution and Eq. (5) the Eddington-Weinberg relation in Eq. (4) immediately leads to
$$t_N^{-2} \sim \Lambda. \tag{6}$$
An equivalent but more physically suggestive form of Eq. (6) may be obtained as follows. The characteristic gravitational energy density $\varepsilon$ within a homogeneous sphere of mass $M$ and radius $R$ is related to the characteristic gravitational acceleration $a$ of the sphere by
$$\varepsilon \sim a \frac{M}{R^2}. \tag{7}$$
Similarly the energy density scale $\varepsilon_N$ characteristic of the gravitational self-energy of the massive quanta of the effective field of the nucleon would be
$$\varepsilon_N \sim \frac{a_N m_N}{\lambda_N^2}. \tag{8}$$
Eq. (7) may be expressed in terms of fundamental parameters as
$$\varepsilon_N \sim \frac{G m_N^6 c^4}{h^4}. \tag{9}$$
If this energy density should correspond to the vacuum density associated with the cosmological constant then it must be that
$$\frac{c^2}{G} \Lambda \sim \frac{G m_N^6 c^4}{h^4}. \tag{10}$$
With the substitution $H_0 \sim \sqrt{\Lambda}$ Eq. (9) is immediately found to be identical to the Eddington-Weinberg relation in Eq. (1). A natural scale for the energy density attributed to the cosmological constant is thus found in the vacuum density associated with the effective field of the nucleon and the holographic conjecture.

This correspondence is rendered more curious by yet another connection. According to the MOND theory, the laws of motion deviate from Newton's Laws for accelerations below some critical value $a_0 \sim 10^{-10} \text{ms}^{-2}$ [4]. Stars within galaxies and galaxies within clusters are typically bound by accelerations of order $a_0$. Remarkably the Hubble acceleration $H_0 c$ is of order the critical acceleration $a_0$ in this epoch. This correlation has not been explained and may constitute another cosmic coincidence. Alternatively, the alignment between the Hubble and critical accelerations in this epoch could be due to an implicit physical scaling
$$a_0 \sim a_N. \tag{11}$$
The existence of such a connection between inertial resistance to acceleration and a field density may be expected if inertia is due to some interaction with a zero-point field [5].